\documentclass{IEEEmce}

\usepackage[colorlinks,urlcolor=blue,linkcolor=blue,citecolor=blue]{hyperref}

\usepackage{hyperref}
\hypersetup{ 
 colorlinks=true, 
 linkcolor=blue, 
 filecolor=blue, 
 citecolor = blue, 
 urlcolor=blue, 
 } 
 
\usepackage{upmath}
\usepackage{graphicx}
\usepackage{subfigure}
\usepackage{array}
\usepackage{booktabs} 
\usepackage{makecell} 
\usepackage{caption} 

\jvol{XX}
\jnum{XX}
\paper{XX}
\jmonth{xxx/xxx}
\publisheddate{DD MM YYYY}
\currentdate{DD MM YYYY}
\jname{IEEE Consumer Electronics Magazine}
\pubyear{YYYY}
\doiinfo{MCE.YYYY.Doi Number}

\begin{document}

\sptitle{Running Head Title} 


\title{Empirical Studies of Large Scale Environment Scanning by Consumer Electronics}


\author{Mengyuan Wang, Yang Liu, Haopeng Wang}


\affil{University of Ottawa}

\author{Haiwei Dong}
\affil{Huawei Canada}

\author{Abdulmotaleb El Saddik}
\affil{University of Ottawa}


\markboth{Running Head Title}{Article Title}

\begin{abstract}
This paper presents an empirical evaluation of the Matterport Pro3, a consumer-grade 3D scanning device, for large-scale environment reconstruction. We conduct detailed scanning (1,099 scanning points) of a six-floor building (17,567 m²) and assess the device's effectiveness, limitations, and performance enhancements in diverse scenarios. Challenges encountered during the scanning are addressed through proposed solutions, while we also explore advanced methods to overcome them more effectively. Comparative analysis with another consumer-grade device (iPhone) highlights the Pro3’s balance between cost-effectiveness and performance. The Matterport Pro3 achieves a denser point cloud with 1,877,324 points compared to the iPhone's 506,961 points and higher alignment accuracy with an RMSE of 0.0118 meters. The cloud-to-cloud (C2C) average distance error between the two point cloud models is 0.0408 meters, with a standard deviation of 0.0715 meters. The study demonstrates the Pro3's ability to generate high-quality 3D models suitable for large-scale applications, leveraging features such as LiDAR and advanced alignment techniques. 
\end{abstract}

\maketitle

\enlargethispage{10pt}


\chapterinitial{The demand for} 3D reconstruction has persisted for many years, and various reconstruction devices are available on the market to meet this need. However, with the rapid development of industries such as digital twins \cite{9984845} and Metaverse \cite{10430223}, the requirements for 3D reconstruction in terms of accuracy and scene scale have significantly increased. While professional-grade reconstruction devices offer high accuracy and stability, their complexity and high price prevent their widespread use. In contrast, consumer-grade scanning devices have gained popularity due to their affordability and user-friendliness. For instance, by 2023, Matterport\footnote{Matterport: 
 \url{https://matterport.com/}} had managed 11.7 million scanned spaces, spanning a total area of 38 billion square feet. Recently, Matterport released the new generation—Matterport Pro3—featuring significant advancements in accuracy, adaptability, and environmental coverage.

Although some previous studies \cite{ geomatics3040030, heritage6090327, Shults} have evaluated consumer-grade devices for 3D reconstruction, most have focused on small- to medium-scale scenarios. Relatively few studies have conducted detailed assessments of consumer devices in large-scale, multi-story building environments. In this paper, we choose the consumer-grade Matterport Pro3 to conduct an empirical study, assessing its effectiveness for large-scale environmental scanning. We aim to determine whether the Matterport Pro3 can provide high-quality 3D reconstructions suitable for general applications, while also analyzing the device's limitations in practical use.

\begin{figure}[htbp]
\vspace*{-10pt}
\centerline{\includegraphics[width=16pc]{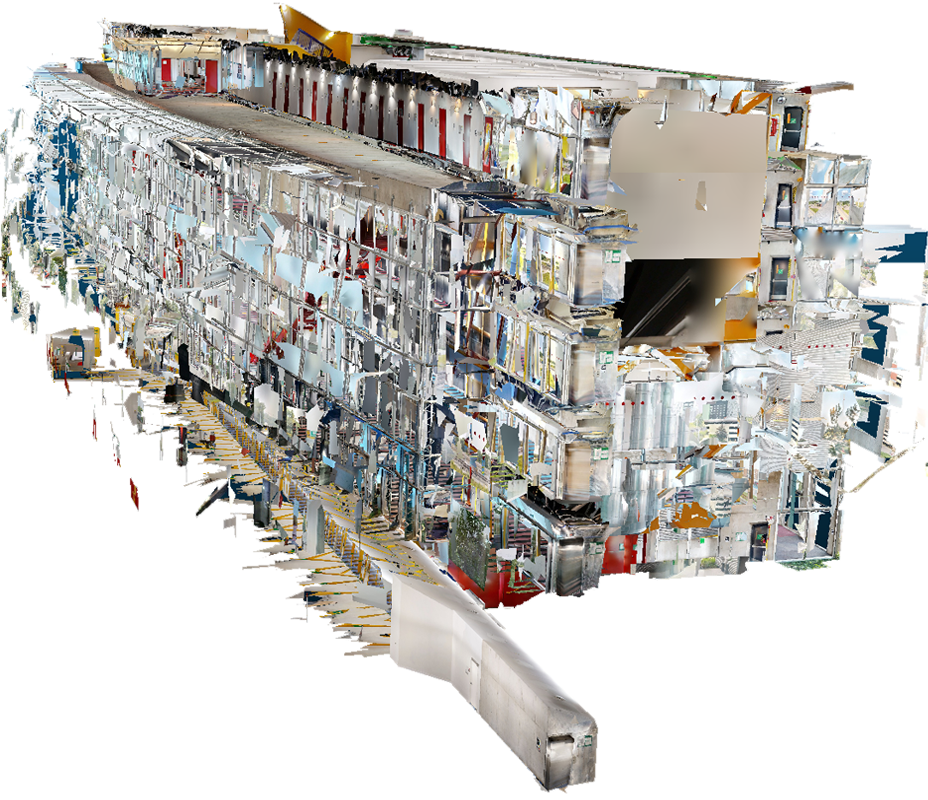}}
\caption{The side view of the reconstructed mesh model illustrates a six-floor structure that includes five above-ground levels and one underground level, encompassing a total area of 17,567 square meters.}
\label{sideview}
\end{figure}

In this study, we use the Matterport Pro3 to scan a six-floor building with a total area of 17,567 square meters, including five above-ground floors and one basement, as shown in Figure \ref{sideview}. The building has a complex design with numerous windows and one wall entirely made of glass. It also features state-of-the-art energy-saving systems such as advanced heating, ventilation, air-conditioning, and ultra-efficient lighting. To ensure comprehensive coverage, we conduct 1,099 scanning points and use the scanned data to construct a high-precision 3D model of the building. 

During the scanning process, several challenges are encountered, including dynamic environment, position mismatch, object misalignment, and model deficiency, which leads to multiple alignment failures. To address these challenges, we conduct a detailed analysis and implement corresponding solutions, such as employing markers for alignment and adjusting scanning settings. As a result of these improvements, we successfully complete a high-precision 3D reconstruction of the building.

\section{RELATED WORK}
In recent years, 3D scanning technology has been widely adopted across diverse fields, including architectural modeling, cultural heritage preservation, virtual reality (VR), and industrial mapping. These technologies utilize depth perception methods such as structured light technology \cite{structured2}, laser radar systems (LiDAR) \cite{lidar}, Time-of-Flight (ToF) sensors \cite{TOF}, and depth-sensing cameras \cite{depthcamera}, integrated into both consumer-grade and professional-grade devices to address diverse performance metrics and application-specific requirements.
\begin{table*}[ht]
\centering
\caption{Comparison of Consumer-Grade and Professional-Grade 3D Scanning Technologies.}
\label{tab:comparison}
\renewcommand{\arraystretch}{0.3} 
\setlength{\tabcolsep}{8pt} 
\begin{tabular}{|>{\centering\arraybackslash}m{1cm}|>{\centering\arraybackslash}m{1.8cm}|>{\centering\arraybackslash}m{3.3cm}|m{3.5cm}|m{3cm}|}
\hline
\textbf{Paper} & \textbf{Device} & \textbf{Sensor \& Accuracy} & \textbf{Summary}                                                                                     & \textbf{Advantages}                                                                                  \\ \hline
Liu et al. \cite{Liu} 
& FARO S-350, Matterport Pro2 
& \makecell[t]{\textbf{TLS:}  Multi-Line LiDAR \\ Accuracy: ±1mm \\ \textbf{Matterport:}  Structured Light \\ Accuracy: ±70mm} 
& Compares TLS and Matterport for heritage building information modeling (HBIM). 
& High precision with TLS; quick, affordable capture with Matterport.                                                                       \\ \hline
Adam et al. \cite{Adam} 
& NavVis VLX 
& \makecell[t]{ Multi-Line LiDAR, \\ RGB Camera \\ Accuracy: ±5mm} 
& Combine an autonomous mobile robot and NavVis VLX to automate the data acquisition process.
& Automated, efficient, and operates during off-hours to avoid disruptions.                                                               \\ \hline
Fiorini et al. \cite{Fiorini} 
& Matterport Pro2, Leica RTC360 
& \makecell[t]{\textbf{Matterport:}  Structured Light \\ Accuracy: ±50mm \\ \textbf{RTC360:}  Multi-Line LiDAR \\ Accuracy: ±3mm} 
& Compare Leica RTC360 with Matterport Pro2, identify issues, and determine optimal solutions.
& Cost-effective and user-friendly (Matterport); highly accurate and versatile (RTC360).                                                  \\ \hline
Shults et al. \cite{Shults} 
& Matterport Pro 3D Camera 
& \makecell[t]{Structured Light \\ Accuracy: ±70mm} 
& The Matterport Pro 3D camera, calibrated with FARO scanner targets, was used to scan the mine interior.
& Fast, low-cost, and integrates well with VR applications (Matterport).                                                                               \\ \hline
Hariffin et al. \cite{Hariffin} 
& Matterport Pro2, FARO Focus3D S350 
& \makecell[t]{\textbf{Matterport:}  Structured Light \\ Accuracy: ±70mm \\ \textbf{FARO:}  Multi-Line LiDAR \\ Accuracy: ±2mm} 
& Examines Matterport's accuracy for strata surveys, showing it is sufficient but less precise compared to TLS.  
& Affordable and easy for complex designs; adequate for strata surveys.                                                                   \\ \hline
Jeftha et al. \cite{lidar}
& iPhone 12 Pro 
& \makecell[t]{Single-Line LiDAR \\ Accuracy: ±6–8cm} 
& Demonstrates the feasibility of iPhone 12 Pro LiDAR for small-scale projects. 
& Affordable, accessible, and effective for non-critical applications.                                                                    \\ \hline
Ours 
& Matterport Pro3 
& \makecell[t]{Single-Line LiDAR, \\ RGB Camera \\ Accuracy: ±20mm @ 10m} 
& Evaluate Matterport Pro3 for a large-scale environment, identify challenges, and propose potential solutions. 
& Affordable and effective for large-scale projects; handles dynamic settings.                                \\ \hline
\end{tabular}
\end{table*}

Professional-grade devices, such as the FARO S-350, Leica RTC360, FARO Focus3D S350, and NavVis VLX, are tailored for high-precision applications, as shown in Table \ref{tab:comparison}. Liu et al. \cite{Liu} demonstrate the FARO S-350’s ±1mm accuracy for heritage building information modeling (HBIM), while Fiorini et al. \cite{Fiorini} highlight the Leica RTC360’s ±3mm accuracy for detailed surveying. Hariffin et al. \cite{Hariffin} evaluate the FARO Focus3D S350, noting its ±2mm accuracy as ideal for strata surveys. Adam et al. \cite{Adam} showcase NavVis VLX’s ±5mm accuracy, highlighting its ability to autonomously scan large, dynamic indoor spaces with high efficiency. However, the high costs of these systems (e.g.,  NavVis VLX \cite{navvis2019} exceeding \$50,000) and their steep learning curves limit their widespread use.

Consumer-grade devices, including the Matterport Pro2, Pro3, and iPhone 12 Pro, prioritize affordability, portability, and ease of use. The Matterport Pro2 achieves ±50–70mm accuracy, making it a cost-effective choice for applications such as cultural heritage documentation \cite{Fiorini} and industrial archaeology \cite{Shults}. Jeftha et al. \cite{lidar} assess the iPhone 12 Pro's ±6–8cm accuracy, making it well-suited for small-scale or non-critical projects. This study extends the literature by evaluating the Matterport Pro3, which bridges the gap between these categories. With ±20mm accuracy at 10 meters, the Pro3 offers a cost-effective solution for large-scale projects.

\section{MATTERPORT PRO3 SYSTEM} 
The Matterport Pro3 is a newly released consumer-grade electronic device designed to create digital twins of large environments. Thanks to advancements in sensor technology and a well-engineered mechanism, the Pro3 captures environments with professional-level precision. Prior to the introduction of the Matterport Pro3, consumer-grade devices are unable to produce high-precision digital twins for expansive environments. Devices like the Matterport Pro2 typically employ structured light to capture depth information for 3D reconstruction.
\begin{figure*}[htbp]
	\centering
	\subfigure[] {\includegraphics[width=.64\textwidth, trim={10 90 0 20},clip]{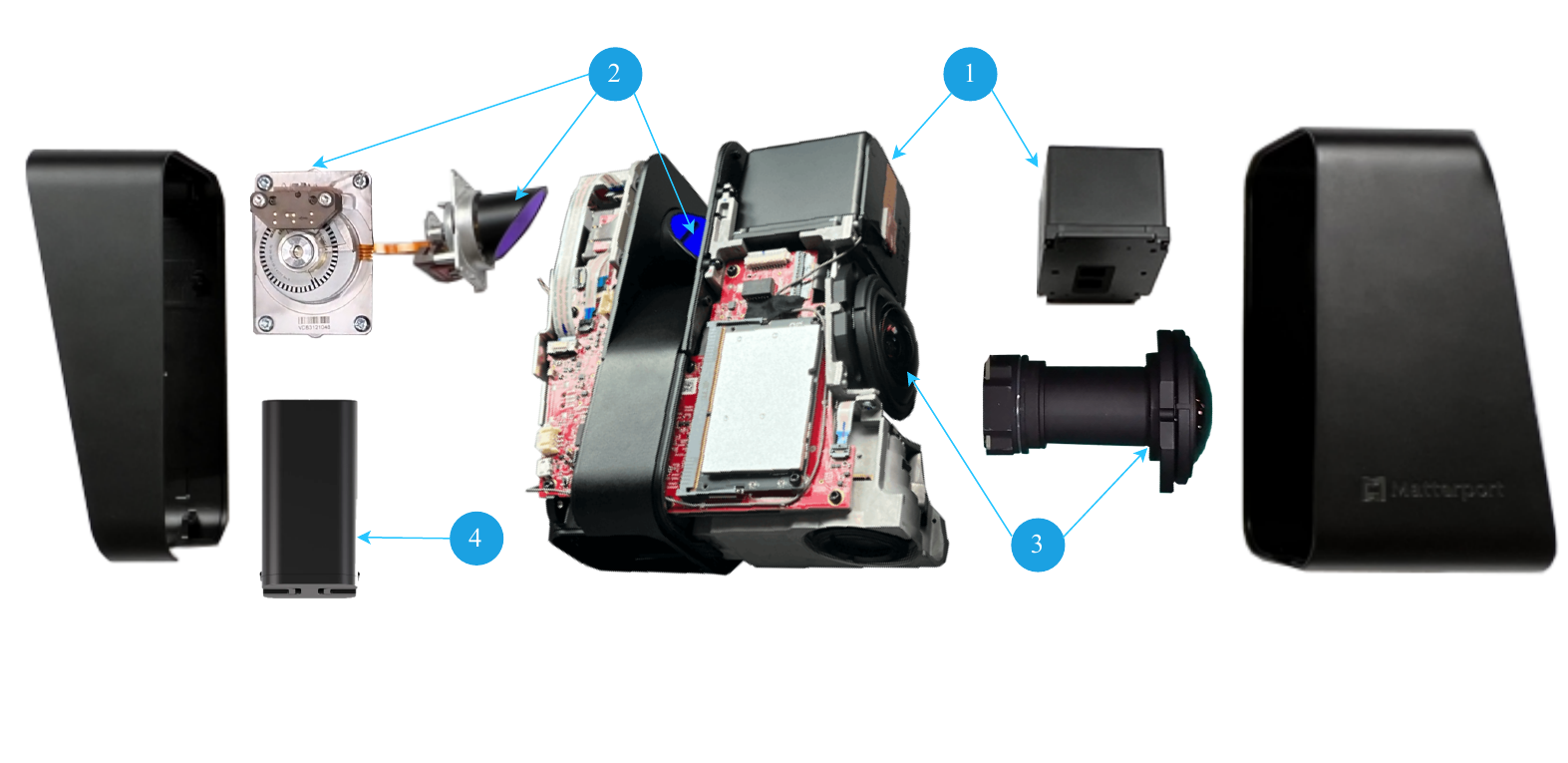}}
	\subfigure[] {\includegraphics[width=.35\textwidth]{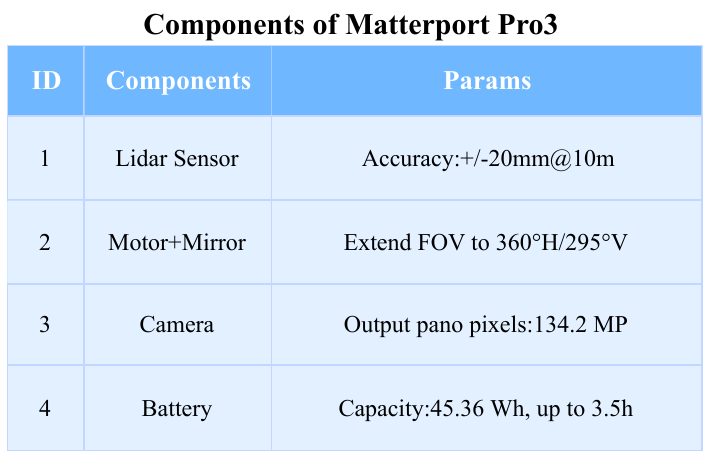}}
	\caption{{Matterport Pro3 Hardware Summary. (a) An exploded view illustrating the primary components of the hardware, each labeled with a blue dot for clarity. (b) A table listing the names and key parameters of the labeled components.}}
	\label{explode_view}
\end{figure*}

In comparison, the Matterport Pro3 takes a more streamlined approach in terms of sensor configuration and mechanical stability to preserve high precision at an affordable price. As illustrated in Figure \ref{explode_view}(a), the Pro3 primarily comprises a LiDAR sensor, a motor with a mirror, a camera, and batteries. The LiDAR sensor meets industry standards, with an error of less than 20 mm at a 10-meter range. Moreover, due to the use of non-repetitive scanning patterns, the single-line LiDAR sensor built in the Pro3 is significantly more affordable than the multi-line LiDAR systems commonly used in industrial applications. Unlike structured light sensors, the LiDAR sensor in the Pro3 overcomes limitations such as poor performance in outdoor environments and inadequate coverage for large areas, providing a strong foundation for large-scale environmental scanning. However, the default field of view (FOV) of the LiDAR sensor is limited. To address this limitation, the Pro3 utilizes a motorized mirror to reflect the LiDAR beams, extending the FOV to a full 360$^{\circ}$ horizontal and 295$^{\circ}$ vertical range, thereby covering a larger area, reducing blind spots, and enhancing data acquisition completeness. As shown in Figure \ref{explode_view}(b), the Pro3 also uses a depth camera that provides an output resolution of 134.2 megapixels, allowing for a more detailed capture of the features and textures of the scene, significantly improving the quality of the reconstruction model. Besides, the Pro3 supports continuous scanning with two removable batteries. Each battery has a capacity of 45.35Wh and can work for 3.5 hours, which makes the Pro3 portable and efficient for indoor applications.  Based on feedback from the six users in this study, the Pro3 shows a short learning curve (less than 30 minutes) and intuitive setup processes such as quick calibration, automatic device recognition, user-friendly app integration, and minimal manual adjustments, allowing operators to begin scanning with ease and efficiency. Additionally, the small size of the Pro3 allows it to fit into tight areas without obstructing movement, while its automated alignment reduces the need for manual adjustments, ensuring accurate scans even in challenging environments with limited space. These characteristics make the Pro3 highly practical for diverse environments.

\section{SCANNING A MULTI-FLOOR BUILDING}
\begin{figure}[htbp]
\centerline{\includegraphics[width=17pc]{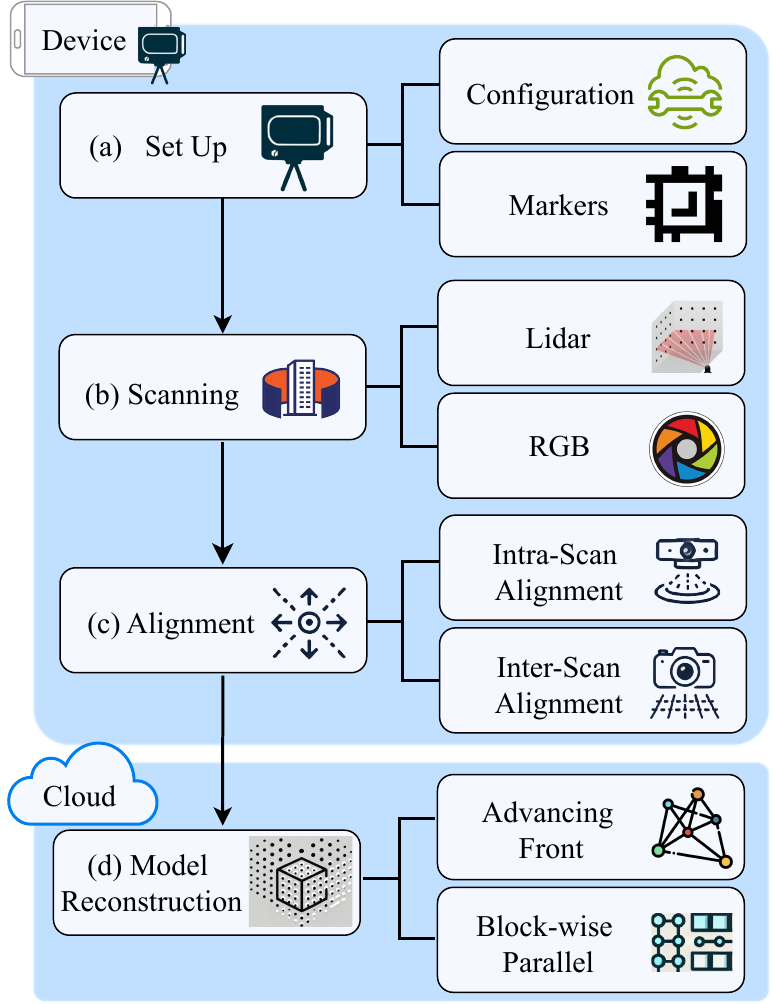}}
\caption{The workflow for creating a high-quality reconstructed mesh for large environments using the Matterport Pro3, which mainly consists of four steps. (a) Setting up the capture devices and markers. (b) Scanning each scene to capture the raw data. (c) Aligning the captured data and scans. (d) Reconstructing the 3D model from the captured data. Steps (a)-(c) are primarily conducted on local devices, while step d is completed in the cloud due to the computational and storage resources required.}
\label{architecture}
\end{figure}

An overview of the Matterport Pro3 workflow for generating a high-quality mesh model based on a large environment is presented in Figure \ref{architecture}. To reconstruct the high-precision 3D model, the process begins with data acquisition using the Pro3 device, which involves setting up the device, performing scans at multiple locations, and aligning these scans to ensure comprehensive coverage. Once the data are captured, it is synchronized with the cloud, where cloud computing resources are utilized to perform high-precision model reconstruction. In this study, we assume that the features identified from multiple viewpoints correspond to the same 3D points in the scene, allowing for accurate feature matching and alignment. In addition, surfaces reflect light uniformly in all directions, simplifying the photometric reconstruction process. The camera is assumed to rotate at a steady rate and the corresponding points in the captured images are expected to maintain consistent color and intensity, facilitating seamless RGB image stitching.

\subsection{Set Up} 
\begin{figure*}[htbp]
\centerline{\includegraphics[width=37pc]{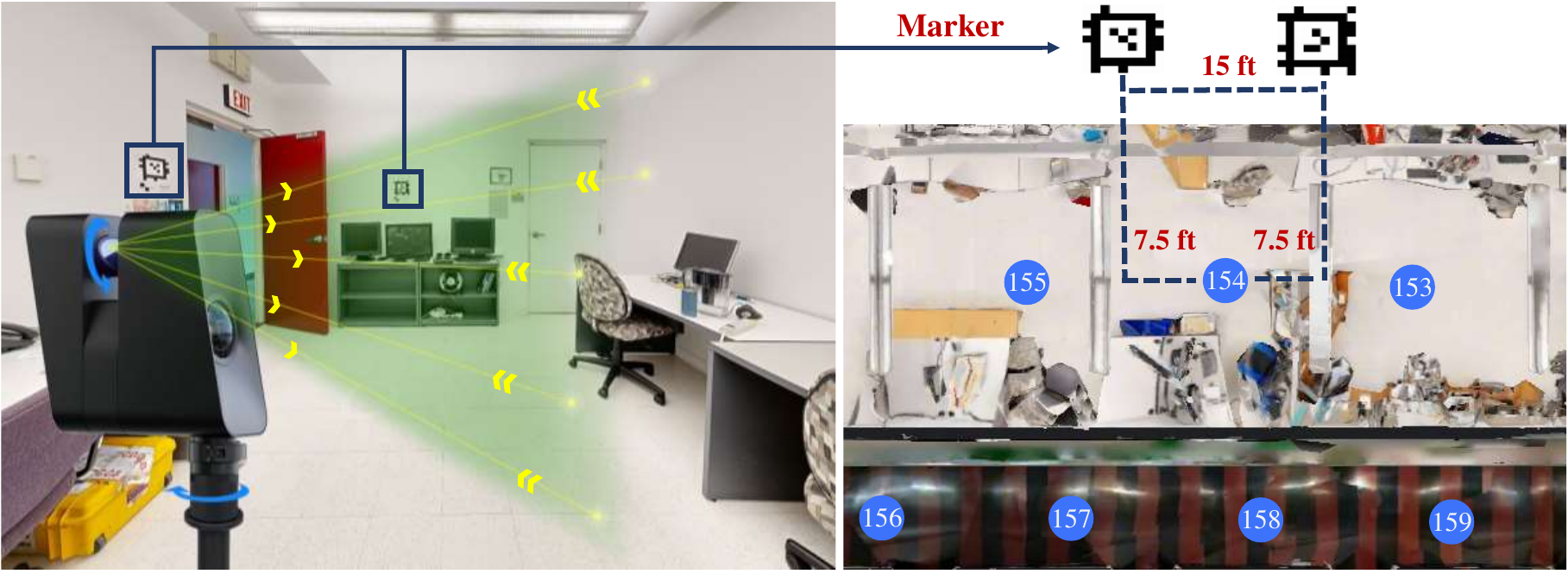}}
\caption{The example setup for scanning. On the left, the Pro3 device is positioned in a room surrounded by markers. During the scanning process, the device rotates according to the indicated blue angle to capture comprehensive 360$^{\circ}$ data. Typically, as shown on the right, the Pro3 is centrally located between two markers, with the blue dots representing the scanning points.}
\label{device}
\end{figure*}
Before starting the scanning process, all major hardware components of the Matterport Pro3, including the LiDAR sensor, motor-driven mirror, and camera, must be calibrated and tested to ensure that they are functioning correctly. For instance, calibrating the gradienter is necessary to maintain the device's level positioning throughout scanning.

Given the large scale and complexity of our building, capturing certain areas can be difficult. For example, if a new scanning area is too similar to the previous one, it can potentially cause misalignment. To address this problem, assisted alignment is used to improve capture efficiency and enhance accuracy. Assisted alignment utilizes visual markers placed at key locations on the scanning site and detected by the Pro3, aiding in the optimization of scan alignment and positioning. However, environments with complex and diverse features often rely less on markers, as environmental landmarks inherently facilitate alignment. As shown in Figure \ref{device}, the dedicated markers are printed out in advance and attached to walls on-site. The markers are posted densely in the building, basically with a distance of about 15 feet between two markers. However, once the markers are placed and scanned, users should not move or reuse them, as this will confuse the identification of the marked positions and affect the overall accuracy and completeness of the scan data.

\subsection{Scanning}
The Matterport Pro3 is positioned between two markers, as shown by the blue dots in Figure \ref{device}. Once scanning begins, the LiDAR sensor emits light waves to capture point cloud data, with the motor driving the mirror to rotate 360$^{\circ}$ to achieve comprehensive coverage. Simultaneously, the Pro3 itself rotates 360$^{\circ}$ to capture high-resolution 2D panoramic images from each scanning position through the camera lens, as depicted in Figure \ref{device}. These images provide color and texture information, which are mapped onto the 3D point cloud captured by the LiDAR sensor, thereby enhancing the visual realism of the final 3D model. Once the current scanning is completed, the Pro3 is moved to the next location, and the building is scanned sequentially. During the Pro3's rotation, the visual markers within the 2D panoramic images are identified. The visual information, including the size and position of the markers, assists in the preliminary placement of each scan, while depth data are subsequently used to fine-tune alignment, ensuring accurate integration of the scanned segments into a cohesive model.
\begin{figure*}[htbp]
\centerline{\includegraphics[width=37pc]{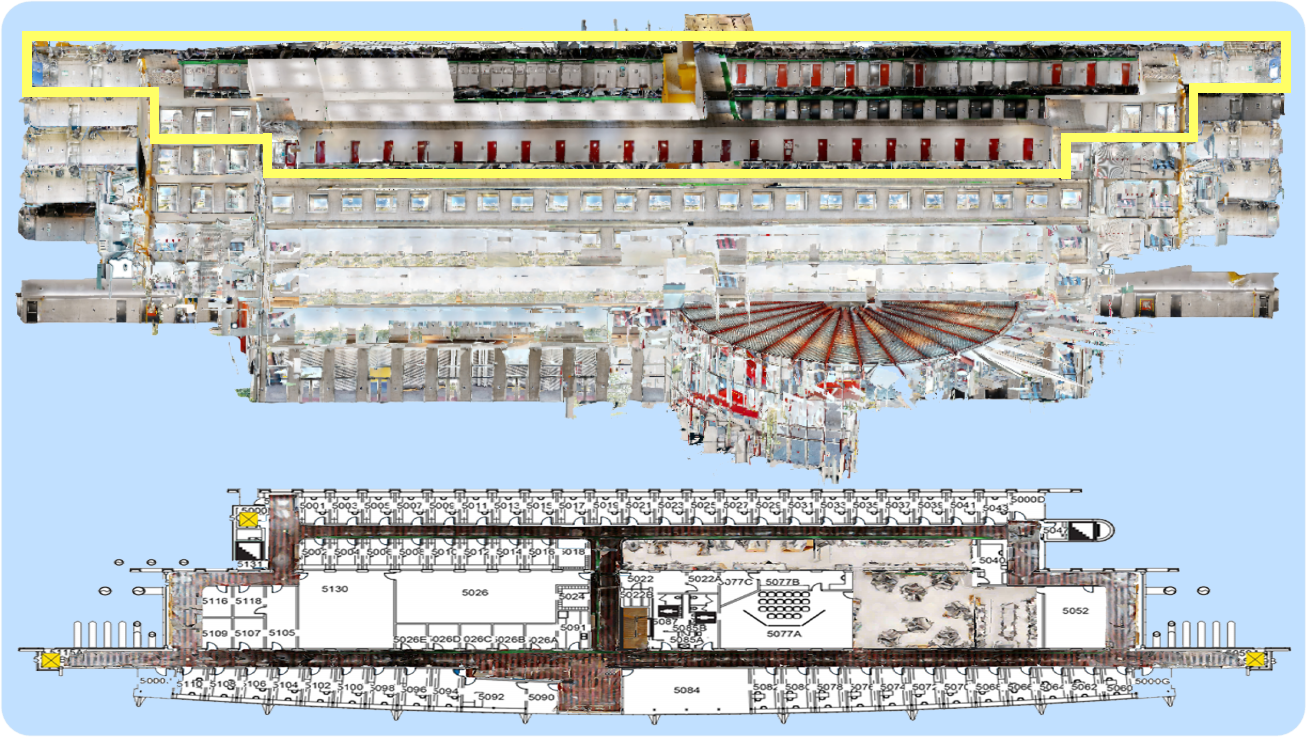}}
\caption{The front view of the reconstructed mesh (upper subfigure) and the top view of the fifth-floor mesh is aligned with the floor plan (lower subfigure).}
\label{outputs}
\end{figure*}

\subsection{Alignment}
Once multiple sensor data have been captured for each scene, data alignment is necessary to fuse the data into a single space. This alignment process comprises two main methods: intra-scan alignment and inter-scan alignment.

Intra-scan alignment is employed to align RGB data with LiDAR data within each individual scan, thereby enhancing the point cloud data with color information. The Matterport device simultaneously collects LiDAR and RGB image data for each scan, ensuring spatial and temporal consistency in the data acquisition process. A rigorous alignment procedure is implemented to integrate the color information into the point cloud. This involves accurately matching LiDAR points with their corresponding pixels in the RGB image. To achieve this, the LiDAR points are projected back onto the image plane using the calibrated camera matrix derived from the device's intrinsic and extrinsic parameters. The calibration ensures that the projection accounts for factors such as lens distortion and spatial offsets between the sensors. Once projected, the color of each point is assigned based on the corresponding pixel value in the RGB image, effectively mapping detailed visual information onto the geometric structure.

\begin{figure*}[htbp]
\centerline{\includegraphics[width=37pc]{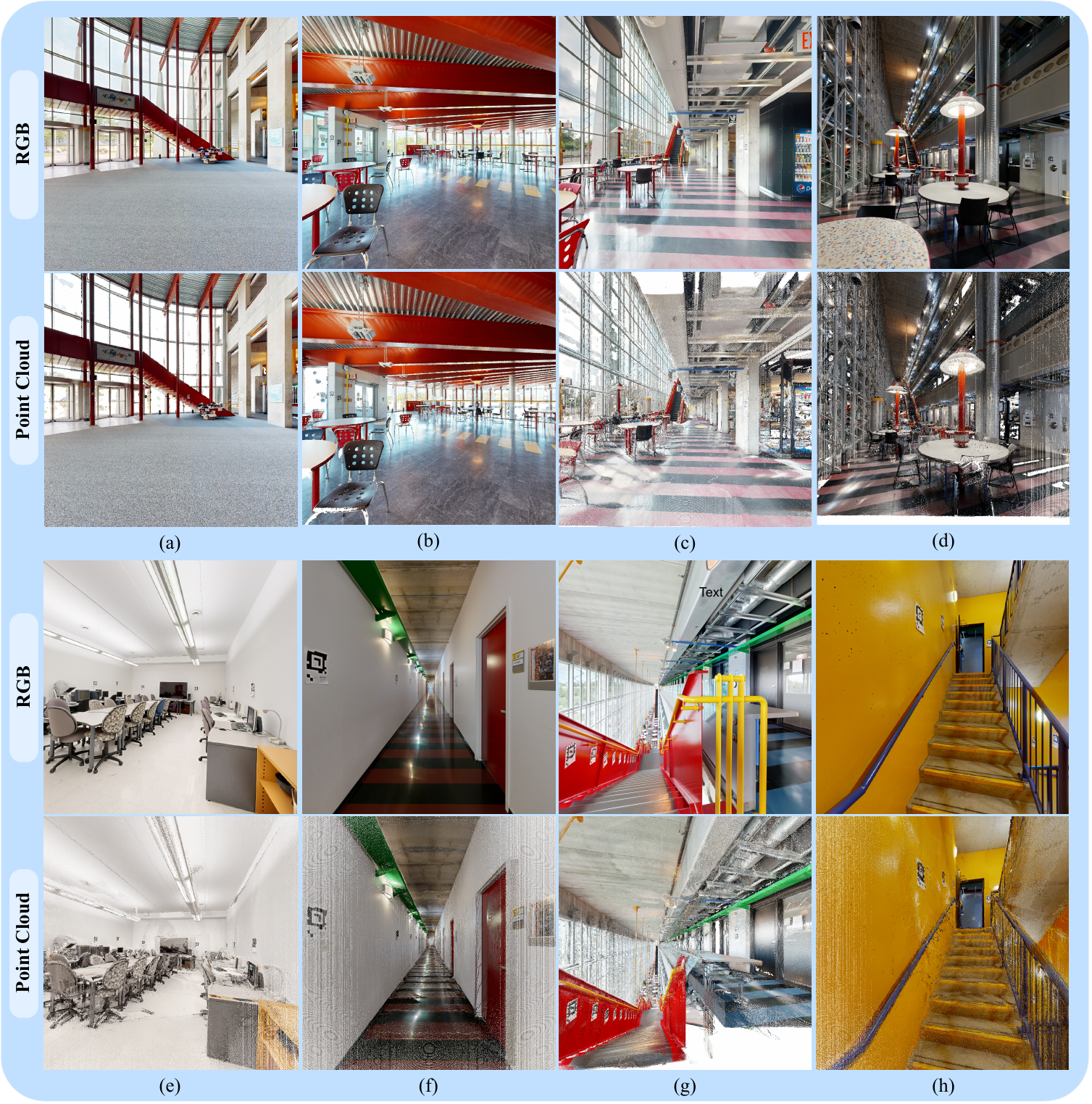}}
\caption{RGB snapshots and corresponding point cloud representations in various scenes. (a) the basement lobby. (b) the first-floor lobby. (c) and (d) are first-floor self-study areas scanned from day to night. (e) a meeting room. (f) a corridor on the 5th floor. (g) a long staircase connecting the first floor to the fifth floor. (h) a staircase exit.}
\label{various scenes}
\end{figure*}

To construct a seamless and accurate 3D environment from multiple scans, a robust inter-scan alignment process is essential, aligning all scans into the same world coordinate system. When a new scan is introduced, it must be precisely aligned with the previously captured data by computing the transformation matrix between the current scan and the existing world coordinate system. This is achieved through an image-based alignment approach, which integrates visual and geometric information to establish accurate correspondences between scans: initially, visually similar image pairs are identified using global image embeddings to measure image similarity across the existing scans. This ensures that only the most relevant image pairs are considered for further processing, reducing computational overhead. Once similar pairs are identified, salient features or key points are extracted using feature detection algorithms, enabling robust identification of corresponding point pairs in 3D between the two scenes. These key points are matched based on their descriptors, ensuring a high degree of reliability in the correspondence. After identifying the matches, the iterative closest point (ICP) algorithm is applied to compute the relative spatial transformation matrix of the new scan. The ICP algorithm refines this transformation by iteratively minimizing the distance between corresponding point pairs in the two point clouds, effectively aligning the scans. Since this process relies heavily on LiDAR data for calculating alignment, the precision and noise characteristics of the LiDAR data at the selected key points play a crucial role in determining the overall accuracy of the intra-scan alignment.

\subsection{Model Reconstruction}
Once the data is captured and the alignment is visually verified, the final model reconstruction is completed in the cloud to leverage the extensive computing resources available. Although Matterport's default cloud service provides a low-quality 3D reconstructed mesh and a high-quality point cloud, the mesh is only suitable for an overview and does not meet the requirements for downstream tasks. Additionally, while the point cloud is useful for measurement purposes, it cannot be directly used in 3D software and platforms. Therefore, we perform 3D reconstruction in the cloud to generate a high-precision model.

To achieve the highest precision in the 3D model, we employ the Advancing Front (AF) surface reconstruction algorithm, which excels in preserving point cloud precision. Unlike other surface reconstruction algorithms, such as Poisson, AF does not assume watertightness, meaning it can accurately model open and complex structures with intricate boundaries that are not fully enclosed. This flexibility makes it ideal for reconstructing real-world objects and environments where gaps or discontinuities may exist. The optimization process in AF involves an iterative advancement of the reconstruction front, dynamically adapting to the spatial distribution of the points. Unlike methods that resample or regularize the point cloud, AF maintains the fidelity of the original data by directly working with the input points, avoiding the introduction of smoothing artifacts or loss of detail. Additionally, AF operates at variable resolutions, which is achieved by adjusting the reconstruction granularity based on the local density of the input point cloud. This adaptive resolution ensures that areas with high point density are captured with fine detail, while sparser regions are reconstructed efficiently, striking a balance between precision and computational efficiency. This optimization process makes the AF algorithm highly effective for large-scale, detailed 3D surface reconstruction tasks.

To efficiently handle large scenes, we adopt a block-wise parallel reconstruction approach. This method divides the scene into smaller, manageable blocks, each processed independently and in parallel, significantly reducing computation time. The process begins with segmenting the scene based on spatial coordinates, ensuring each block contains a manageable number of points. Each block is then reconstructed using the AF algorithm, capturing fine details accurately. Overlapping regions between adjacent blocks are used to align and merge them, minimizing artifacts and ensuring continuity. Once all blocks are reconstructed, they are merged to form the complete scene, resolving any boundary discrepancies to produce a coherent final mesh.

\subsection{Reconstructed 3D Model}
The pipeline's output is a comprehensive suite of data that includes a high-precision 3D reconstructed model, an aligned point cloud, RGB images for each scan, and their corresponding camera poses. 

Specifically, the upper subfigure of Figure \ref{outputs} represents the front view of the reconstructed building model, which comprises six floors—five above ground and one underground. The grid accurately reflects the spatial arrangement of rooms, elevators, and windows, highlighting the Matterport Pro3's capability to capture complex building elements. Notably, the model captures a significant amount of glass from various angles, including a prominent six-story glass wall, showcasing the complexity and detail achieved during the scanning process. These results underscore the thoroughness of the reconstruction and the intricacy of the captured data. To qualitatively assess the reconstructed 3D mesh’s performance, the lower subfigure of Figure 5 shows the top view of the fifth floor aligned with the official building floor plan. Since the top view of the fifth floor is able to consistently match the floor plan, it means that the reconstructed model does not have notable translation or rotation drift and the accumulated error for the reconstructed model is acceptable based on the visual check.

Various scenes and their corresponding point cloud representations are shown in Figure \ref{various scenes}. Open areas such as the basement and first-floor halls (Figure \ref{various scenes}(a) and (b)), and study areas with a glass wall (Figure \ref{various scenes}(c) and (d)) are transparent and well-lit. In sharp contrast, the long and narrow corridor (Figure \ref{various scenes}(f)) is surrounded by concrete walls on both sides and receives limited light. In addition, the staircase area also shows the diversity of architectural design, with narrow stairs (Figure \ref{various scenes}(h)) and long stairs (Figure \ref{various scenes}(g)) highlighting the vertical extension and ductility of the interior space of the building, respectively. These diverse scenarios are common in modern urban architecture. Meanwhile, the building was scanned under different lighting conditions from day to night, which further underscores the study's applicability to similar environments and supports the generalization of its findings.

\section{PERFORMANCE ANALYSIS}
 To evaluate the performance of consumer-grade scanning devices, we compare the Matterport Pro3 with an iPhone 3D scanner using the same real-world environment. The Matterport Pro3 generates a dense point cloud with 1,877,324 points, whereas the iPhone 3D scanner produces a sparse point cloud with 506,961 points. The scanned datasets are processed and compared using CloudCompare\footnote{CloudCompare: 
 \url{https://www.danielgm.net/cc/}}. The process begins by aligning the Matterport point cloud to the same reference coordinate system as the iPhone point cloud. The ICP algorithm was then utilized to enhance the manual alignment procedure. Following alignment, the cloud-to-cloud (C2C) distance was computed to quantify the geometric differences between corresponding points in the two point cloud models.
\begin{figure*}[htbp]
	\centering
	\subfigure[] {\includegraphics[width=.55\textwidth]{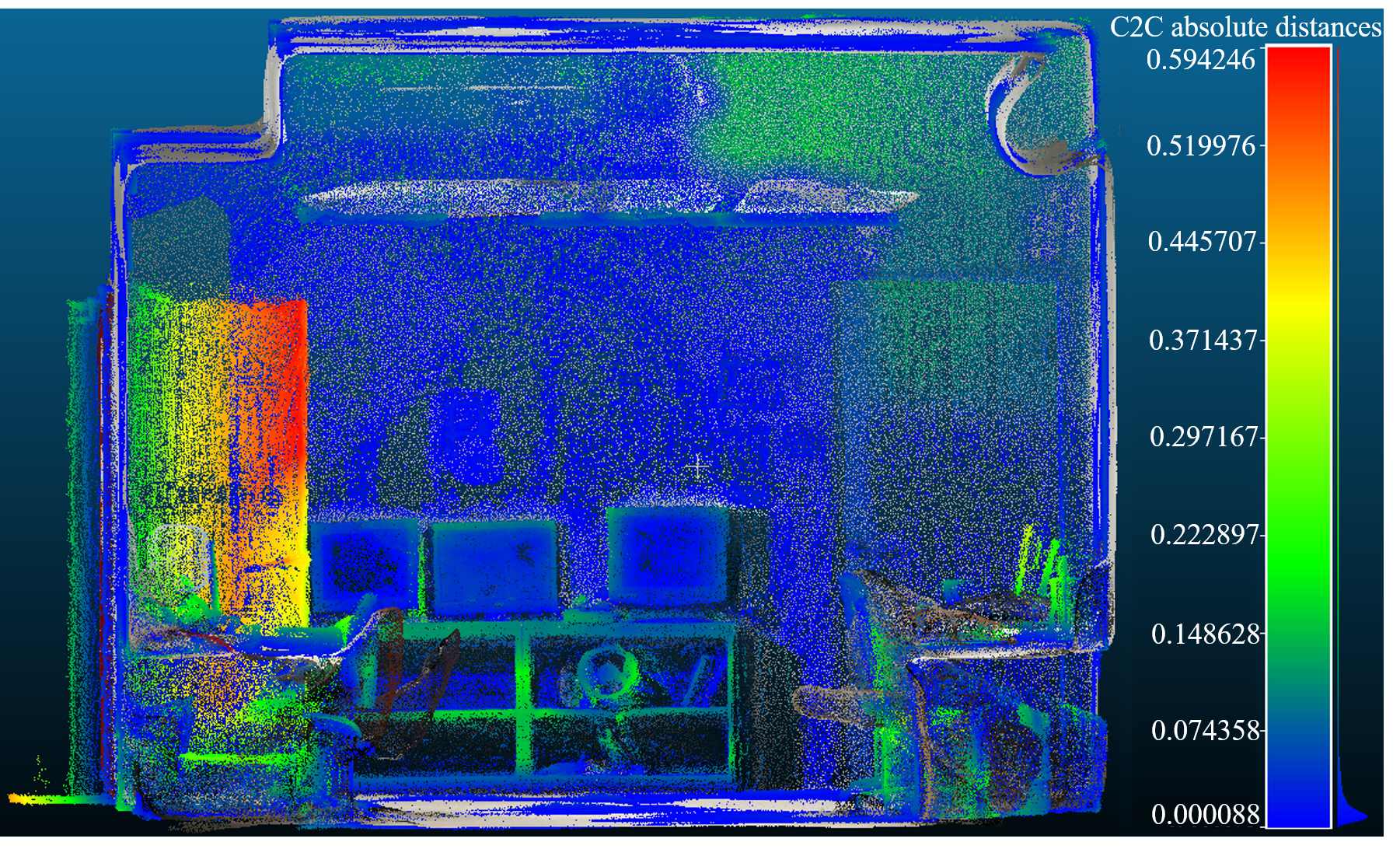}}
	\subfigure[] {\includegraphics[width=.44\textwidth, trim={10 10 10 20},clip]{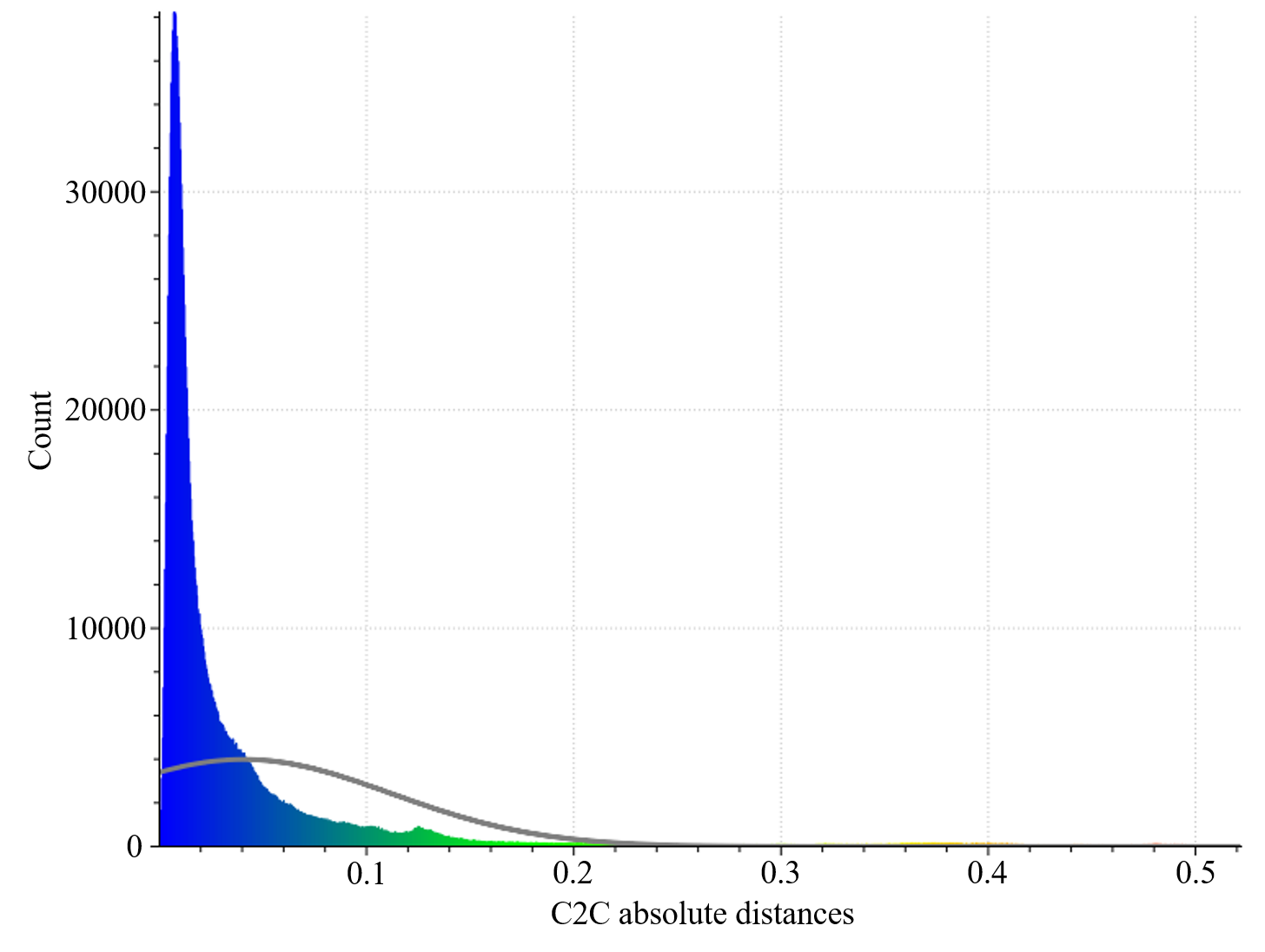}}
	\caption{Comparison of the scanned 3D models between the Matterport Pro3 and iPhone 3D scanner. (a) Heatmap of cloud-to-cloud distance difference between models. (Reference: iPhone 3D scanner point cloud; Aligned: Matterport Pro3 point cloud). (b) Gaussian distribution of absolute distance errors across the scene.}
	\label{point cloud comparison difference}
\end{figure*}

A heat map in Figure \ref{point cloud comparison difference}(a) shows the different distances between the two point clouds captured by the Matterport Pro3 and the iPhone 3D scanner: regions with larger distances are shown in red, while those with smaller distances are displayed in blue. In this analysis, the maximum cloud-to-cloud distance is 0.52 meters. In particular, the distance between the doors of the room ranges from 0.32 and 0.52 meters. The Gaussian distribution graph in Figure \ref{point cloud comparison difference}(b) further shows a mean difference of 0.0408 meters between points from the Pro3 and the iPhone, with a standard deviation of 0.0715 meters, indicating that most point errors are small and concentrated. Additionally, the root mean square error (RMSE) of the Matterport model is 0.0118 meters, reflecting a high global consistency. In contrast, the low structural similarity index (SSI) of 0.0025 reflects significant differences in point density and geometric representation, which underscores the Matterport Pro3's superior detail capture compared to the iPhone's lower-resolution output. Therefore, the Matterport Pro3 outperforms the iPhone 3D scanner with its higher point density and lower alignment error, making it ideal for accurate and high-resolution 3D modeling, while the iPhone scanner is suited to quick, low-cost tasks in simpler environments.

\section{CHALLENGES AND DISCUSSIONS}

\begin{figure}[htbp]
\vspace*{-10pt}
\centerline{\includegraphics[width=17pc]{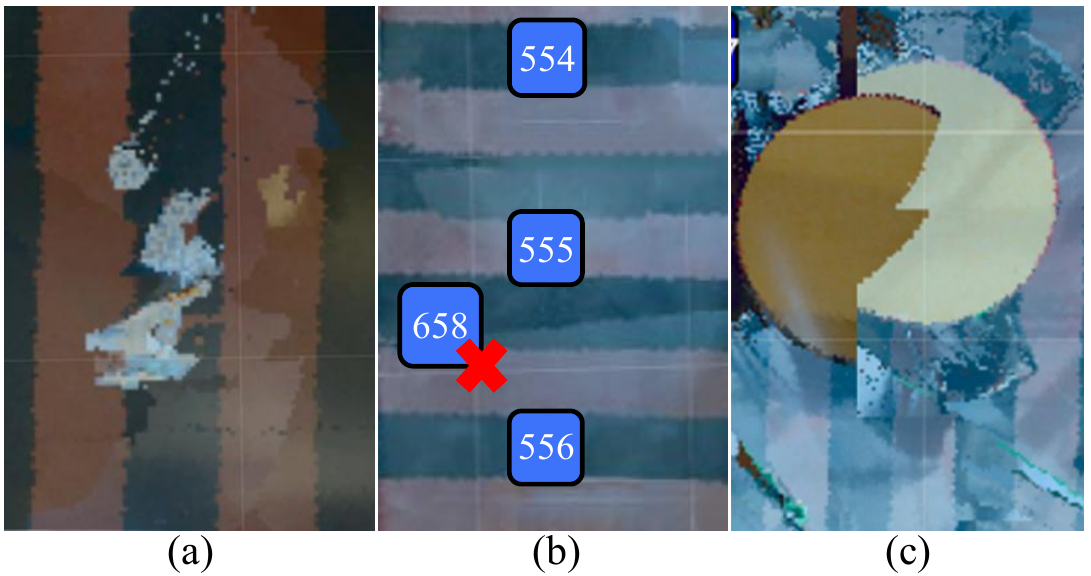}}
\caption{The challenges encountered during scanning include (a) human ghosting; (b) position mismatch; and (c) object misalignment.}
\label{challenges}
\end{figure}

In this study, we successfully conduct detailed 3D scanning of a building characterized by diverse and complex architectural features. The scanning process presented several challenges, such as alignment issues caused by glass reflections, dynamic elements, and cumulative inherent errors, as illustrated in Figure \ref{challenges}, we identify these challenges and find optimal solutions. Since our proposed methods primarily focus on environmental adjustments and post-processing, the solutions demonstrate broad applicability for other devices with similar capabilities.
The following sections will detail the most common challenges we faced and the potential solutions we identified to address them.

\begin{itemize}
    \item \textbf{Dynamic Environment} 
     Frequent movement of people during scanning can lead to challenges such as ghosting and the unintended inclusion of individuals in the 3D model, as illustrated in Figure \ref{challenges}(a). Additionally, minor shifts in furniture, like tables and chairs, can introduce inconsistencies and alignment errors in the final model. These issues primarily arise because the model directly utilizes the noisy point cloud captured by LiDAR data for moving objects. Although the Matterport system provides an automatic facial blurring feature, it does not entirely remove individuals from the captured RGB images, nor does it offer post-processing for LiDAR data to address this. To mitigate these issues, each scan undergoes meticulous review to exclude individuals and moving objects, thereby minimizing misalignment in dynamic environments. Furthermore, conducting multiple scans from the same point and employing advanced data fusion techniques with RGB images and LiDAR data can effectively eliminate noisy data when dynamic environments are unavoidable. Unfortunately, this capability is not currently supported by the Matterport system.

    \item  \textbf{Position Mismatch} 
    Another common issue is the misalignment or incorrect positioning of scan points. As depicted in Figure \ref{challenges}(b), scan point 658 was erroneously matched to the opposite side of the floor. This typically occurs when the retrieval algorithm fails to identify the correct reference scan in the existing map due to missing visual features. In the example shown, the narrow, long corridor with repetitive patterns led the algorithm to incorrectly retrieve scan point 555, resulting in a positional error. Variations in lighting conditions and reflections can further complicate the alignment process. To address this, incorporating additional visual markers can significantly reduce such errors. In scenarios where adding visual markers is not feasible, localization priors become essential. For instance, professional devices like NavVis utilize Wi-Fi fingerprints to narrow the retrieval search space. Implementing automatic localization priors in consumer devices could enhance their adaptability to diverse, large environments.

    \item \textbf{Object Misalignment} 
    During scanning, we observed a noticeable misalignment of certain static objects in the final reconstructed model at scan overlap areas, as exemplified in Figure \ref{challenges}(c), where a single yellow desk is split into two parts. This issue primarily arises from small rotational and translational errors in nearly every camera pose estimation, which accumulates with each subsequent scan, leading to visible misalignment in overlap areas. To address this, we refine the camera pose before reconstruction. Specifically, we extract dense visual features using SuperPoint and apply loop closure and global optimization to minimize accumulated errors. This approach effectively eliminates object misalignment for static objects. Incorporating default support for global alignment in custom devices could significantly reduce complexity for customers, making the technology more accessible and user-friendly. 

    \item \textbf{Model Deficiency}
    The final model exhibits deficiencies in sheltered or obstructed areas, such as the spaces between fire hydrants and walls, which cannot be fully captured during scanning. Additionally, due to the presence of glass walls and extensive windows in the building, the interaction between laser beams and glass surfaces can lead to phenomena such as penetration, reflection, refraction, and scattering. These interactions lead to several issues. Firstly, a portion of the laser beam penetrates the glass and continues propagating without returning to the sensor, resulting in missing or incomplete data in the affected areas. Secondly, the laser beams can be reflected multiple times within the glass wall and windows because of their thickness, which can cause structural duplication in the final model. Lastly, imperfections on the glass surface, such as scratches and contamination, can scatter or diffract the laser beams, causing deviations from the expected trajectory and further degrading the accuracy of the scan. Solving these issues requires significant post-processing, including correcting these errors and filling in data gaps to improve the accuracy and completeness of the final model.
   
\end{itemize} 
\section{LIMITATION AND FUTURE WORK}
This study evaluates the Matterport Pro3 across a large-scale building environment, identifying its challenges and proposing general solutions. In the future, we will also validate the Pro3's performance in diverse environments to enhance the generalizability of the solutions. While we evaluate the Matterport Pro3's cost-efficiency and ease of use, it cannot fully replace professional-grade devices that provide superior accuracy and environmental adaptability. However, consumer devices can serve as viable alternatives. Improving consumer-grade devices offers a fast and cost-effective solution to large-scale reconstruction challenges, delivering immediate performance gains. For example, integrating advanced hardware, such as high-resolution LiDAR sensors, can capture reliable data in limited light or reflective conditions. Depth cameras can enhance point cloud density in dynamic or cluttered environments. Developing new technologies requires longer cycles and higher investments, but such advancements can overcome fundamental limitations and enable transformative progress. A combined approach balances immediate needs with future innovation, ensuring practicality and long-term development.

While practical solutions have been suggested to address the challenges identified in this study, the use of AI-based or automated advanced algorithms for tackling specific issues, such as human ghosting, glass reflections, and incomplete data, has not been explored. Since this was beyond the scope of our study, future work will focus on advanced technologies to improve scanning accuracy and efficiency in dynamic and complex environments, including real-time filtering and adaptive strategies during scanning, improved alignment methods, enhanced pose estimation for global consistency, automated marker placement, and advanced post-processing techniques to address artifacts and occlusion. These advancements aim to streamline the scanning process, reduce manual effort, and improve the accuracy and reliability of 3D reconstruction. For instance, AI-driven algorithms can enable real-time error detection and correction during scanning by dynamically adapting to changes, filtering noise, and optimizing alignment. Automated rescanning can identify and target areas with insufficient data, while advanced pose refinement algorithms ensure consistent alignment, minimize manual intervention, and enhance reconstruction accuracy and efficiency.

\section{CONCLUSION}
In this study, we conduct an empirical evaluation of the Matterport Pro3 for detailed scanning of a six-floor building spanning 17,567 square meters. We provide a detailed overview of the device's key hardware components, present a comprehensive scanning workflow, and explore advanced algorithms designed to optimize the scanning process and enhance the reconstruction of high-quality 3D models. Additionally, we perform a comparative experiment with another consumer-grade device to evaluate the relative performance in terms of point cloud density and alignment accuracy. The results highlight the capability of the Pro3 to capture complex and large-scale environments with high accuracy. Several common challenges are encountered, such as alignment issues and dynamic environmental factors, for which we propose practical solutions and suggest potential device enhancements. These recommendations aim to streamline the scanning process, enhance device performance, and support future large-scale environment reconstruction efforts.


\begin{IEEEbiography}{Mengyuan Wang}
is a PhD Candidate at the University of Ottawa. Her research interests include multimedia, Metaverse, and Digital Twin. Contact her at mwang259@uottawa.ca.
\end{IEEEbiography}

\begin{IEEEbiography}{Yang Liu}
is a PhD Candidate at the University of Ottawa and a member of IEEE. His research interests include artificial intelligence and multimedia. Contact him at yliu344@uottawa.ca.
\end{IEEEbiography}

\begin{IEEEbiography}{Haopeng Wang} is a Postdoctoral Fellow with the School of Electrical Engineering and Computer Science, University of Ottawa, Ottawa, ON, Canada. His research interests include multimedia, extended reality, and artificial intelligence. Wang received his Ph.D. degree from the University of Ottawa. Contact him at hwang266@uottawa.ca.
\end{IEEEbiography}

\begin{IEEEbiography}{Haiwei Dong} is a Principal Researcher and Director at Huawei Canada, and an Adjunct Professor at the University of Ottawa. His research interests include artificial intelligence, multimedia, Metaverse, and robotics. Dong received his Ph.D. degree from Kobe University, Kobe, Japan. He is a Senior Member of the IEEE and ACM. Contact him at haiwei.dong@huawei.com.
\end{IEEEbiography}

\begin{IEEEbiography}{Abdulmotaleb El Saddik} is a Distinguished University Professor with the University of Ottawa, Ottawa, ON, Canada. His research interests include multimodal interactions with sensory information in smart cities. He is a Fellow of Royal Society of Canada, a Fellow of IEEE, an ACM Distinguished Scientist and a Fellow of the Engineering Institute of Canada and the Canadian Academy of Engineers. Contact him at elsaddik@uottawa.ca.
\end{IEEEbiography}


\begin{thebibliography}{00}
\bibitem{9984845} H. Dong and Y. Liu, ``Metaverse meets consumer electronics,'' {\it IEEE Consum. Electron. Mag.}, vol. 12, no. 3, pp. 17--19, 2023, doi: 10.1109/MCE.2022.3229180.

\bibitem{10430223} H. Wang, R. Martinez-Velazquez, H. Dong, and A. El Saddik, ``Experimental studies of metaverse streaming,'' {\it IEEE Consum. Electron. Mag.}, vol. 14, no. 1, pp. 26--36, 2024.

\bibitem{geomatics3040030} C. Askar and H. Sternberg, ``Use of smartphone lidar technology for low-cost 3D building documentation with iPhone 13 Pro: A comparative analysis of mobile scanning applications,'' {\it Geomatics}, vol. 3, no. 4, pp. 563--579, 2023, doi: 10.3390/geomatics3040030.

\bibitem{heritage6090327} R. Shults, E. Levin, Z. Aukazhiyeva, K. Pavelka, N. Kulichenko, N. Kalabaev, M. Sagyndyk, and N. Akhmetova, ``A study of the accuracy of a 3D indoor camera for industrial archaeology applications,'' {\it Heritage}, vol. 6, no. 9, pp. 6240--6267, 2023, doi: 10.3390/heritage6090327.

\bibitem{structured2} G. Lin, H. Zhang, S. Xie, J. Luo, Z. Li, and Y. Wang, ``Research on point cloud structure detection of manhole cover based on structured light camera,'' {\it Electron.}, vol. 13, no. 7, pp. 1--14, 2024, doi: 10.3390/electronics13071226.

\bibitem{lidar} K. J. Jeftha and M. Shoko, ``Mobile phone based laser scanning as a low-cost alternative for multidisciplinary data collection,'' {\it S. Afr. J. Sci.}, vol. 120, no. 11/12, pp. 1--7, 2024, doi: 10.17159/sajs.2024/15437.

\bibitem{TOF} Z. Sun, W. Ye, J. Xiong, G. Choe, J. Wang, S. Suochen, and R. Ranjan, ``Consistent direct time-of-flight video depth super-resolution,'' {\it Proc. 2023 IEEE/CVF Conf. Comput. Vision Pattern Recognit.}, pp. 5075--5085, 2023, doi: 10.1109/CVPR52729.2023.00491.

\bibitem{depthcamera} P. Gholami and R. Xiao, ``AutoDepthNet: High frame rate depth map reconstruction using commodity depth and RGB cameras,'' {\it arXiv Prepr.}, May 2023, doi: 10.48550/arXiv.2305.14731.

\bibitem{Liu} J. Liu, D. Willkens, C. López, L. Cortés-Meseguer, J. L. García-Valldecabres, P. A. Escudero, and S. Alathamneh, ``Comparative analysis of point clouds acquired from a TLS survey and a 3D virtual tour for HBIM development,'' {\it Int. Arch. Photogramm. Remote Sens. Spatial Inf. Sci.}, vol. XLVIII-M-2-2023, pp. 959--968, 2023, doi: 10.5194/isprs-archives-XLVIII-M-2-2023-959-2023.

\bibitem{Fiorini} G. Fiorini, M. A. Tini, F. Montelli, and G. Bitelli, ``Comparison of two technologies in 3D surveying of real estate assets and cultural heritage,'' {\it Proc. MetroArchaeo 2023}, pp. 435--440, 2023, doi: 10.21014/tc4-ARC-2023.083.

\bibitem{Hariffin} N. A. B. Hariffin, M. H. B. Razali, L. C. Luh, S. A. Sulaiman, and M. B. M. Hashim, ``The suitability of Matterport for building parcel dimension survey,'' {\it Proc. 2023 IEEE 13th Int. Conf. Syst. Eng. Technol.}, pp. 239--244, 2023, doi: 10.1109/ICSET59111.2023.10295134.

\bibitem{Adam} M. G. Adam, M. Piccolrovazzi, A. Dalloul, C. Werner, and E. Steinbach, ``Continuous and autonomous digital twinning of large-scale dynamic indoor environments,'' {\it Proc. IEEE 19th Int. Conf. Autom. Sci. Eng.}, pp. 1--6, 2023, doi: 10.1109/CASE56687.2023.10260595.

\bibitem{navvis2019} NavVis US Inc., NavVis. [Online]. Available: {https://www.navvis.com/}

\bibitem{Shults} R. Shults, E. Levin, R. Habibi, S. Shenoy, O. Honcheruk, T. Hart, and Z. An, ``Capability of Matterport 3D camera for industrial archaeology sites inventory,'' {\it Int. Arch. Photogramm. Remote Sens. Spatial Inf. Sci.}, vol. XLII-2/W11, pp. 1059--1064, 2019, doi: 10.5194/isprs-archives-XLII-2-W11-1059-2019.

\end{thebibliography}
\end{document}